\pdfoutput=1
\documentclass[12pt, epsfig]{article}
\usepackage{amsmath,amsfonts,amssymb,amsthm,amstext,amscd,eucal,graphicx,xcolor}
\usepackage[all]{xy}
\usepackage{dsfont,epiolmec}
\usepackage{amsmath}
\usepackage{slashed,ccaption,cite}
\usepackage{mathrsfs, calligra}

\usepackage[unicode=true,
 bookmarks=false,
 breaklinks=false,pdfborder={0 0 1},backref=false,colorlinks=true]
 {hyperref}
\hypersetup{
 pdfauthor={Daniel Grumiller, Shahin Sheikh-Jabbari },
 linkcolor=blue,urlcolor=magenta,filecolor=green,citecolor=red,pdfstartview=FitV,bookmarksopen=true}

\makeatletter \@addtoreset{equation}{section}

\makeatletter\renewcommand\section{\@startsection {section}{1}{\z@}%
                                   {-3.5ex \@plus -1ex \@minus -.2ex}%nn
                                   {2.3ex \@plus.2ex}%
                                   {\normalfont\large\bfseries}}
\renewcommand\subsection{\@startsection{subsection}{2}{\z@}%
                                     {-3.25ex\@plus -1ex \@minus -.2ex}%
                                     {1.5ex \@plus .2ex}%
                                     {\normalfont\bfseries}}

\renewcommand{\baselinestretch}{1.2}

\makeatletter
\DeclareFontFamily{OMX}{MnSymbolE}{}
\DeclareSymbolFont{MnLargeSymbols}{OMX}{MnSymbolE}{m}{n}
\SetSymbolFont{MnLargeSymbols}{bold}{OMX}{MnSymbolE}{b}{n}
\DeclareFontShape{OMX}{MnSymbolE}{m}{n}{
    <-6>  MnSymbolE5
   <6-7>  MnSymbolE6
   <7-8>  MnSymbolE7
   <8-9>  MnSymbolE8
   <9-10> MnSymbolE9
  <10-12> MnSymbolE10
  <12->   MnSymbolE12
}{}
\DeclareFontShape{OMX}{MnSymbolE}{b}{n}{
    <-6>  MnSymbolE-Bold5
   <6-7>  MnSymbolE-Bold6
   <7-8>  MnSymbolE-Bold7
   <8-9>  MnSymbolE-Bold8
   <9-10> MnSymbolE-Bold9
  <10-12> MnSymbolE-Bold10
  <12->   MnSymbolE-Bold12
}{}

\let\llangle\@undefined
\let\rrangle\@undefined
\DeclareMathDelimiter{\llangle}{\mathopen}%
                     {MnLargeSymbols}{'164}{MnLargeSymbols}{'164}
\DeclareMathDelimiter{\rrangle}{\mathclose}%
                     {MnLargeSymbols}{'171}{MnLargeSymbols}{'171}
\makeatletter
\newcommand{\be}{\begin{equation}}
\newcommand{\ee}{\end{equation}}
\newcommand{\bea}{\begin{eqnarray}}
\newcommand{\eea}{\end{eqnarray}}
\newcommand{\bse}{\begin{subequations}}
\newcommand{\ese}{\end{subequations}}
\newcommand{\beqa}{\begin{eqnarray}}
\newcommand{\eeqa}{\end{eqnarray}}
\newcommand{\beqar}{\begin{eqnarray*}}
\newcommand{\eeqar}{\end{eqnarray*}}
\newcommand{\bi}{\begin{itemize}}
\newcommand{\ei}{\end{itemize}}
\newcommand{\bn}{\begin{enumerate}}
\newcommand{\en}{\end{enumerate}}
\newcommand{\ba}{\begin{array}}
\newcommand{\ea}{\end{array}}
\newcommand{\bc}{\begin{center}}
\newcommand{\ec}{\end{center}}

\newcommand{\eq}[2]{\begin{equation} #1 \label{#2} \end{equation}}

\DeclareMathOperator{\extdm}{d}
\newcommand{\extd}{\extdm \!}

\definecolor{darkgreen}{rgb}{0,0.3,0}
\definecolor{darkblue}{rgb}{0,0,0.3}
\definecolor{darkred}{rgb}{0.7,0,0}

\newcommand{\old}[1]{}%{\sout{#1}}

\begin{document}
\renewcommand{\baselinestretch}{1.2}  %Line spacing

\begin{titlepage}

\begin{flushright}\vspace{-3cm}
{
TUW--18--02, IPM/P-18/011\\
March 28, 2018 }\end{flushright}

\newcommand{\mytitle}{Membrane Paradigm from Near Horizon Soft Hair}

\vspace*{2truecm}

\begin{center}
\Large{\bf{\hspace*{-.5cm} \mytitle}}\\

\bigskip
%\bigskip

\large{\bf{D.~Grumiller\footnote{corresponding author; e-mail:~\href{grumil@hep.itp.tuwien.ac.at}{grumil@hep.itp.tuwien.ac.at}}$^{; a}$ and M.M.~Sheikh-Jabbari\footnote{e-mail:~\href{jabbari@theory.ipm.ac.ir}{jabbari@theory.ipm.ac.ir}}$^{; b}$}}
\\

\normalsize
\bigskip

{$^a$ \it Institute for Theoretical Physics, TU Wien, Wiedner Hauptstr.~8, A-1040 Vienna, Austria
 }
\\
{$^b$ \it School of Physics, Institute for Research in Fundamental
Sciences (IPM),\\ P.O.Box 19395-5531, Tehran, Iran}\\

\end{center}
\setcounter{footnote}{0}

\bigskip

\begin{abstract}
The membrane paradigm posits that black hole microstates are dynamical
degrees of freedom associated with a physical membrane  vanishingly close to the black hole's event horizon. The soft hair paradigm postulates that black
holes can be equipped with zero-energy charges associated with residual
diffeomorphisms that label near horizon degrees of freedom. In this essay
we argue that the latter paradigm implies the former.
More specifically, we exploit suitable near horizon boundary conditions
that lead to an algebra of ``soft hair charges'' containing infinite copies
of the Heisenberg algebra, associated with area-preserving shear
deformations of black hole horizons. We employ the
near horizon soft hair and its Heisenberg algebra to provide a formulation
of the membrane paradigm and show how it accounts for black hole entropy.
\end{abstract}

\vskip1cm
\centerline{\textit{This essay is prepared for the Gravity Research Foundation, 2018 Awards for Essays on Gravitation}}

\end{titlepage}

\newpage

\vspace*{0.1truecm}

\paragraph{
{\LARGE{\textbf{\calligra{B}}}}
lack holes} are spacetimes characterized by two causally disconnected regions, separated by an event horizon. For stationary generic black holes, on which we mainly focus in this essay, the horizon is a bifurcate null surface, with past and future horizons intersecting on the bifurcation surface, a compact spacelike surface of co-dimension two. Therefore, constant time slices of black hole spacetimes are divided into inside and outside regions. The horizon is classically a semi-permeable membrane through which things fall in but do not come out.

Black holes behave like thermal states with a large entropy, as pioneered by Bekenstein \cite{Bekenstein:1973ur}, at a certain temperature, first calculated by Hawking \cite{Hawking:1974rv}. %,Hawking:1974sw}. 
According to Einstein's equivalence principle an outside observer should be able to give a complete description of physics with the information available to her, without the need to know the inside horizon information. Therefore, one expects to be able to associate black hole thermodynamics to a region outside the horizon, infinitesimally close to it, the so-called stretched horizon \cite{tHooft:1990fkf}. %,Susskind:1993if}.

The membrane paradigm \cite{Price:1986yy} %,Thorne:1986iy} 
was originally proposed to give a simple classical picture for black holes and their (thermo)dynamics, particularly from the outside observer's viewpoint, without recrossing into intricate and subtle points of quantum field theory on curved backgrounds containing causally disconnected regions. The paradigm is based on replacing the region inside the stretched horizon by a thin, classically radiating, membrane that embodies thermodynamical properties of the black hole. This membrane  has a tension, can be electrically charged and conducting, has finite entropy and temperature, but cannot conduct heat. The interaction of the stretched horizon with the external universe is described in terms of the Navier--Stokes equation associated with the membrane, Ohm's law, a tidal-force equation, and the first and second laws of thermodynamics.

Studying quantum field theory on a black hole background yields Hawking radiation. This confirms the thermodynamical picture discussed above and shows that stationary black holes, like any thermodynamical system in equilibrium, emit black-body radiation (at the Hawking temperature). It is hence natural to pose the following question: can the membrane paradigm be extended semi-classically?

To answer this question one needs to convert the membrane paradigm into a specific membrane model, which particularly accounts for the black hole entropy and its associated microstates. Different aspects of this issue were addressed in light of black hole complementarity \cite{tHooft:1990fkf}, %,Susskind:1993if}, 
which was challenged a few years ago through the firewall controversy \cite{Almheiri:2012rt}. %,Almheiri:2013hfa}. 
While there are other approaches for formulating the membrane paradigm, e.g.~\cite{Son:2007vk},  %,Bhattacharyya:2008jc,Iqbal:2008by,Faulkner:2010jy,Eling:2016xlx}, 
in this essay we revisit and motivate it from a novel viewpoint. We start from the concept of near-horizon soft hair and the associated near horizon symmetries, which we shall spell out explicitly for generic black holes in four spacetime dimensions. As we shall see, our soft hair analysis unexpectedly leads us back to the membrane paradigm. Among other things, our approach provides us with a statistical mechanical setup to account for black hole entropy within the membrane picture.

\paragraph{Near-horizon soft hair.} %is going to be our main technical tool, so let us step back for a moment and consider its pre-history. In the early 1960's Bondi, van der Burg, Metzner and Sachs (BMS) \cite{Bondi:1962,Sachs:1962} discovered for the specific example of asymptotically flat spacetimes that in gravity there could be an infinite tower of non-trivial conserved charges associated with specific coordinate transformations, calling for a refinement of the equivalence principle. Configurations related by such transformations are considered as physically distinct, and in certain circumstances can correspond to zero energy excitations, i.e., soft modes. This research line attracted a lot of attention in recent years, especially after the seminal work of Hawking, Perry and Strominger (HPS) \cite{Hawking:2016msc,Hawking:2016sgy} who introduced the notion of \emph{soft hair}. The name ``soft hair'' alludes to the famous statement by John Wheeler ``black hole have no hair'', referring to the fact that black hole solutions are uniquely specified by a handful of parameters like mass, angular momentum and electric charge; while classical black holes have no hair in Wheeler's sense, HPS argued that they can be dressed by zero energy (=soft) excitations.

Hawking, Perry and Strominger (HPS) \cite{Hawking:2016msc} %,Hawking:2016sgy} 
introduced about two years ago the notion of \emph{soft hair}, alluding to the famous statement by John Wheeler ``black holes have no hair''. While classical black holes have no hair in Wheeler's sense (meaning that they are uniquely specified by a handful of parameters like mass, angular momentum and electric charge), HPS argued that black holes can be dressed by zero energy (=soft) excitations.

Near horizon soft hair emerges from the imposition of suitable near horizon boundary conditions, like the ones discovered in \cite{Donnay:2015abr}. %,Donnay:2016ejv}. 
The near horizon boundary conditions proposed in \cite{Afshar:2016wfy} %,Afshar:2016kjj} 
led to infinite copies of the Heisenberg algebra as near horizon symmetries, albeit only in three spacetime dimensions, and to a specific semi-classical proposal for black hole microstates \cite{Afshar:2016uax}. %,Sheikh-Jabbari:2016npa,Afshar:2017okz,Hajian:2017mrf}. 
Building on ongoing work with Perez and Troncoso \cite{essay2018} we generalize now the boundary conditions of \cite{Afshar:2016wfy} to four dimensions with the aim to exhibit again infinite copies of the Heisenberg algebra as near horizon symmetries. Expanding the metric around the horizon at $\rho=0$ yields
\eq{
\extd s^2 = -\kappa^2\rho^2\extd t^2+\extd\rho^2+ 2\rho N_a \extd\rho \extd x^a+ \Omega_{ab} \extd x^a \extd x^b +{\cal O}(\rho^2)\qquad\qquad a,b=1,2,
}{eq:bc1}
where $\kappa$ is (constant and fixed) surface gravity and $\Omega_{ab}$ is the metric on the bifurcate horizon ${\cal H}$. Metric fluctuations of order
\eq{
\delta g_{\rho a}={\cal O}(\rho)\,,\quad \delta g_{ab} = {\cal O}(1) \qquad \Rightarrow \qquad N_a = {\cal O}(1) = \delta  \Omega_{ab},
}{eq:bc2}
turn out to generate near horizon soft hair, while all other fluctuations either are determined from these soft hair excitations or correspond to small diffeomorphisms. 

The full charge analysis will appear elsewhere \cite{essay2018}; here we just quote its main results: there are three sets of non-trivial diffeomorphisms and associated near horizon charges: 1.~near horizon supertranslations, 2.~area preserving shear deformations (APSDs), and 3.~area preserving twist deformations. Since we are ultimately interested in black hole microstates we freeze the latter, as twist deformations are not expected to affect the microstates and their number. The supertranslations are generated by angle-dependent time-translations, $\xi^t(\theta,\,\varphi)$, and the APSDs by the gradient part of diffeomorphisms of the 2-sphere, $\xi^a_{\textrm{\tiny grad}}(\theta,\,\varphi)$. (The twist deformations would correspond to the curl part.) The associated near horizon charges
\eq{
Q[\xi^t,\,\xi^a_{\textrm{\tiny grad}}] = \int_{{\cal H}}\!\extd^2x\,\big[\xi^t\,{\cal P} + \xi^a_{\textrm{\tiny grad}}\,\partial_a \Phi \big]
}{eq:nh1}
contain the state-dependent functions ($\Omega:=\det\Omega_{ab}$)
\eq{
{\cal P} = \frac{\sqrt{\Omega}}{8\pi G}\qquad\qquad \partial_a \Phi = \frac{\Omega_{ab}\,\pi^{\rho b}}{8\pi G\,\sqrt{\Omega}}\Big|_{\textrm{\tiny grad}},
}{eq:nh2}
where $G$ is Newton's constant, $\pi^{\rho{b}}$ is the canonical momentum associated with the metric component $g_{\rho{b}}$ and the subscript `grad' means that we take only the gradient contribution (and not the curl part) on the right hand side of the right equality \eqref{eq:nh2}. In contrast to the boundary conditions of \cite{Donnay:2015abr} we use densitized superrotation parameters $\xi^a_{\textrm{\tiny grad}}$, which is a small but significant change reminiscent of the transformation between standard and Ashtekar variables \cite{Ashtekar:1986yd}.

\paragraph{Near horizon Heisenberg algebra and entropy.}
A straightforward canonical analysis reveals the following Poisson brackets between the supertranslation charges ${\cal P}$ and APSD-charges $\Phi$
\begin{align}
    \{\Phi(x),\,{\cal P}(y)\} &= \frac{1}{8\pi G}\,\delta^{(2)}(x-y) \label{eq:key} \\
    \{{\cal P}(x),\,{\cal P}(y)\} &= 0, \qquad\qquad \{\Phi(x),\,\Phi(y)\} = {\cal B}, 
\end{align}
where the explicit form of ${\cal B}$ is not needed here. Note that ${\cal B}$ vanishes trivially for non-rotating black holes (like Schwarzschild) and for black holes with flat horizons (like toroidal Kerr-anti-de~Sitter) \cite{essay2018}. %, but not for Kerr. %For simplicity we consider henceforth the case ${\cal B}=0$, though we expect that all our main considerations generalize to non-vanishing $\cal B$. 
%\dnote{to be checked if removable even for Kerr}

The commutation relation \eqref{eq:key} is a key result and means that the charge associated with supertranslations is the Heisenberg conjugate of the charge associated with APSDs. This result has profound physical consequences and will lead to a formulation of the membrane paradigm, accounting also for the black hole entropy. 
To discuss some of the physical consequences note that the algebra \eqref{eq:key} is nothing but the Heisenberg algebra, where the role of Planck's constant $h$ is played by a quarter of the inverse of Newton's constant.
\eq{
h \simeq \frac{1}{4G}
}{eq:key2}
We stress that the identification \eqref{eq:key2} is not an input but rather a result emerging from our near horizon analysis.

The near-horizon Hamiltonian $H=Q[\xi^t=\kappa,\,\xi^a=0]=\kappa\,\int_{\cal{H}}\extd^2x\,{\cal P}=:\kappa\,{\cal P}_0$ has vanishing Poisson brackets with all near horizon symmetries, which proves the softness property. Its variation 
\be\label{NH-H}
\delta H = \kappa\,\delta {\cal P}_0 = T\,\delta S
\ee
establishes the near-horizon first law of thermodynamics, thereby recovering the Bekenstein--Hawking entropy (using $T=2\pi/\kappa$)
\eq{
S=2\pi{\cal P}_0=\frac{1}{4G}\int_{{\cal H}} \extd^2x\ \sqrt{\Omega} = \frac{\textrm{Area}}{4G}\,.
}{eq:BH}
The entropy formula \eqref{eq:BH} is in accordance with previous near horizon results \cite{Donnay:2015abr,Afshar:2016wfy} and expected from the (Iyer--)Wald analysis \cite{Wald:1993nt}. %,Iyer:1994ys}. 
Before addressing how our near horizon results above relate to the membrane paradigm we consider briefly an example.

\paragraph{The Kerr black hole} with mass $M=(r_++r_-)/2$ and rotation parameter $a=\sqrt{r_+r_-}$, but without any soft hair excitations, yields the near horizon charges ($\theta$ is the polar angle on a unit $S^2$)
\eq{
{\cal P}_{\textrm{\tiny Kerr}} = \frac{M(M+\sqrt{M^2-a^2})\,\sin\theta}{4\pi G}\qquad\qquad \Phi_{\textrm{\tiny Kerr}} = 0\,.
}{eq:angelinajolie}
The condition that the area preserving twist deformations are fixed implies the scaling relations $\delta\textrm{Area}/\textrm{Area}=2\delta{r_+/r_+}=2\delta{r_-/r_-}$. Thus, if one keeps the area fixed, $\delta\textrm{Area}=0$, the absence of twist deformations implies that mass $M$ and Kerr parameter $a$ are fixed as well. The remaining soft hair excitations (supertranslations and APSDs) do not affect mass or Kerr parameter and thus have a chance to label microstates of a given Kerr black hole.

\paragraph{Reloading membrane paradigm with near-horizon soft hair.} The dynamics of a membrane is goverened by an action invariant under diffeomorphisms that preserve the volume it sweeps in spacetime. For a membrane wrapping the horizon ${\cal H}$ this includes the APSDs. The dynamics of a codimension one membrane is then described by a single degree of freedom ${\cal P}$ and its Heisenberg conjugate $\Phi$. This perfectly matches with our near-horizon symmetry analysis above, once we identify $1/(4G)$ as the membrane tension, as suggested by the identification \eqref{eq:key2}. In other words, \eqref{eq:key} may be viewed as the constant time Poisson bracket, and semi-classically as corresponding commutator, of the degrees of freedom of the membrane theory. The simplest membrane action compatible with these considerations is
\eq{
I_{\textrm{\tiny membrane}}=\frac{1}{4G}\,\int \extd t_E \extd x^2 \sqrt{h_{ij}}
}{eq:m} 
where $h_{ij}$ is the induced metric. Fixing static gauge on the membrane, the action \eqref{eq:m} evaluated in the Euclidean section for the membrane wrapping the stretched horizon (integrating Euclidean time $t_E$ over a thermal circle of unit radius) yields the Bekenstein--Hawking entropy \eqref{eq:BH}. 
Thus, the membrane naturally accounts for %the area law and 
the black hole entropy given by the logarithm of the partition function derived from the membrane on-shell action \eqref{eq:m}. 

\paragraph{Summary and outlook.} Even if one did not know about the membrane paradigm, our near-horizon soft hair analysis not only indirectly implies it and provides a venue for its precise formulation but may also enable us to account for black hole microstates from semi-classical first principles, conceptually along the lines of a recent proposal \cite{Afshar:2016uax}. %,Sheikh-Jabbari:2016npa,Afshar:2017okz,Hajian:2017mrf}. 
Here we only presented a semi-classical analysis that accounts for the entropy, without explicitly specifying the microstates. For the latter one needs to carry out quantization of the membrane action, which may be performed %in light-cone gauge 
along the lines of \cite{deWit:1988wri}. %,Hoppe:2002km}. 
Like in three dimensions \cite{Afshar:2016uax}, a key step will be a controlled cutoff on the soft hair spectrum that selects a precise subset of soft hair excitations as black hole microstates. Our hope is that the relation to the membrane paradigm indicated above can provide such a cutoff.

%Since our computations are based on near-horizon analyses they work for black holes with any asymptotic behavior. Moreover, while here we focused on four-dimensional black holes, our analysis readily generalizes to generic (non-extremal) black holes and even cosmologies in any dimension \cite{essay2018}. In accord with expectations, the black hole entropy only knows about the gravitational part of the theory and is not directly affected by matter. This, among other things, addresses the species problem \cite{Jacobson:1994iw} %,Susskind:1994sm} 
%of the usual membrane paradigm. 

Finally, it would be excellent to establish a connection between our analysis and holographic entanglement entropy \cite{Ryu:2006ef}, shedding further light on black hole complementarity \cite{tHooft:1990fkf} %,Susskind:1993if} 
and the resolution of the firewall issue \cite{Almheiri:2012rt}. %,Almheiri:2013hfa}. 
Indeed, the boundary conditions \eqref{eq:bc1}, \eqref{eq:bc2} eliminate singularities on the horizon and the whole soft hair spectrum is compatible with regularity, suggesting the absence of firewalls. %\dnote{there are many loose ends in the previous three paragraphs; perhaps ok for an essay, but maybe not optimal?}

% in case you are wondering: the symbol below is the Epi-Olmec symbol for "star"
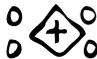
\begin{center}
{\LARGE{\EOstar}}
\end{center}

%\newpage

\noindent\textbf{Acknowledgement}

We dedicate this essay to the memory of Stephen Hawking and Joe Polchinski.
We thank our collaborators, in particular Hamid Afshar, Alfredo Perez and Ricado Troncoso, for many discussions over the years and for their role in the development of the picture and ideas presented in this essay. M.M.~Sh-J is partially supported by the grants from ICTP NT-04, INSF grant No 950124. D.G.~is supported by the Austrian Science Fund (FWF), projects P 27182-N27 and P 28751-N27. We acknowledge an Iran-Austria IMPULSE project grant.

%\bibliographystyle{fullsort}
%\bibliography{review}

\begin{thebibliography}{10}

\bibitem{Bekenstein:1973ur}
J.~D. Bekenstein, ``Black holes and entropy,'' {\em Phys. Rev.} {\bf D7} (1973)
2333--2346.
%%CITATION = PHRVA,D7,2333;%%.

\bibitem{Hawking:1974rv}
S.~W. Hawking, ``Black hole explosions,'' {\em Nature} {\bf 248} (1974)
30--31.
%%CITATION = NATUA,248,30;%%.
%
%\bibitem{Hawking:1974sw}
``Particle creation by black holes,'' {\em Commun. Math. Phys.}
  {\bf 43} (1975)
199--220.
%%CITATION = CMPHA,43,199;%%.

\bibitem{tHooft:1990fkf}
G.~'t~Hooft, ``{The black hole interpretation of string theory},'' {\em Nucl.
  Phys.} {\bf B335} (1990)
138--154.
%%CITATION = NUPHA,B335,138;%%.

%\bibitem{Susskind:1993if}
L.~Susskind, L.~Thorlacius, and J.~Uglum, ``{The Stretched horizon and black
  hole complementarity},'' {\em Phys.Rev.} {\bf D48} (1993) 3743--3761,
\href{http://www.arXiv.org/abs/hep-th/9306069}{{\tt hep-th/9306069}}.
%%CITATION = HEP-TH/9306069;%%.

\bibitem{Price:1986yy}
R.~H. Price and K.~S. Thorne, ``{Membrane Viewpoint on Black Holes: Properties
  and Evolution of the Stretched Horizon},'' {\em Phys. Rev.} {\bf D33} (1986)
915--941.
%%CITATION = PHRVA,D33,915;%%.

%\bibitem{Thorne:1986iy}
K.~S. Thorne, R.~Price, and D.~Macdonald,
{\em Black holes: The membrane paradigm}, Yale University Press 1986.
%%CITATION = ISBN-9780300037708 ETC.;%%.

\bibitem{Almheiri:2012rt}
A.~Almheiri, D.~Marolf, J.~Polchinski, and J.~Sully, ``{Black Holes:
  Complementarity or Firewalls?},'' {\em JHEP} {\bf 1302} (2013) 062,
\href{http://www.arXiv.org/abs/1207.3123}{{\tt arXiv:1207.3123}}.
%%CITATION = ARXIV:1207.3123;%%.

%\bibitem{Almheiri:2013hfa}
A.~Almheiri, D.~Marolf, J.~Polchinski, D.~Stanford, and J.~Sully, ``{An
  Apologia for Firewalls},'' {\em JHEP} {\bf 1309} (2013) 018,
\href{http://www.arXiv.org/abs/1304.6483}{{\tt arXiv:1304.6483}}.
%%CITATION = ARXIV:1304.6483;%%.

\bibitem{Son:2007vk}
D.~T. Son and A.~O. Starinets, ``{Viscosity, Black Holes, and Quantum Field
  Theory},'' {\em Ann. Rev. Nucl. Part. Sci.} {\bf 57} (2007) 95--118,
\href{http://www.arXiv.org/abs/0704.0240}{{\tt arXiv:0704.0240}}.
%%CITATION = ARXIV:0704.0240;%%.

%\bibitem{Bhattacharyya:2008jc}
S.~Bhattacharyya, V.~E. Hubeny, S.~Minwalla, and M.~Rangamani, ``{Nonlinear
  Fluid Dynamics from Gravity},'' {\em JHEP} {\bf 02} (2008) 045,
\href{http://www.arXiv.org/abs/0712.2456}{{\tt arXiv:0712.2456}}.
%%CITATION = ARXIV:0712.2456;%%.

%\bibitem{Iqbal:2008by}
N.~Iqbal and H.~Liu, ``{Universality of the hydrodynamic limit in AdS/CFT and
  the membrane paradigm},'' {\em Phys.Rev.} {\bf D79} (2009) 025023,
\href{http://www.arXiv.org/abs/0809.3808}{{\tt arXiv:0809.3808}}.
%%CITATION = ARXIV:0809.3808;%%.

%\bibitem{Faulkner:2010jy}
T.~Faulkner, H.~Liu, and M.~Rangamani, ``{Integrating out geometry: Holographic
  Wilsonian RG and the membrane paradigm},'' {\em JHEP} {\bf 08} (2011) 051,
\href{http://www.arXiv.org/abs/1010.4036}{{\tt arXiv:1010.4036}}.
%%CITATION = ARXIV:1010.4036;%%.

%\bibitem{Eling:2016xlx}
C.~Eling and Y.~Oz, ``{On the Membrane Paradigm and Spontaneous Breaking of
  Horizon BMS Symmetries},'' {\em JHEP} {\bf 07} (2016) 065,
\href{http://www.arXiv.org/abs/1605.00183}{{\tt arXiv:1605.00183}}.
%%CITATION = ARXIV:1605.00183;%%.

\bibitem{Hawking:2016msc}
S.~W. Hawking, M.~J. Perry, and A.~Strominger, ``{Soft Hair on Black Holes},''
  {\em Phys. Rev. Lett.} {\bf 116} (2016), no.~23, 231301,
\href{http://www.arXiv.org/abs/1601.00921}{{\tt arXiv:1601.00921}}.
%%CITATION = ARXIV:1601.00921;%%.
%
%\bibitem{Hawking:2016sgy}
%S.~W. Hawking, M.~J. Perry, and A.~Strominger, 
``{Superrotation Charge and Supertranslation Hair on Black Holes},'' {\em JHEP} {\bf 05} (2017) 161,
\href{http://www.arXiv.org/abs/1611.09175}{{\tt arXiv:1611.09175}}.
%%CITATION = ARXIV:1611.09175;%%.

\bibitem{Donnay:2015abr}
L.~Donnay, G.~Giribet, H.~A. Gonzalez, and M.~Pino, ``{Supertranslations and
  Superrotations at the Black Hole Horizon},'' {\em Phys. Rev. Lett.} {\bf 116}
  (2016), no.~9, 091101,
\href{http://www.arXiv.org/abs/1511.08687}{{\tt arXiv:1511.08687}}.
%%CITATION = ARXIV:1511.08687;%%.
%
%\bibitem{Donnay:2016ejv}
%L.~Donnay, G.~Giribet, H.~A. González, and M.~Pino, 
``{Extended Symmetries at
  the Black Hole Horizon},'' {\em JHEP} {\bf 09} (2016) 100,
\href{http://www.arXiv.org/abs/1607.05703}{{\tt arXiv:1607.05703}}.
%%CITATION = ARXIV:1607.05703;%%.

\bibitem{Afshar:2016wfy}
H.~Afshar, S.~Detournay, D.~Grumiller, W.~Merbis, A.~Perez, D.~Tempo, and
  R.~Troncoso, ``{Soft Heisenberg hair on black holes in three dimensions},''
  {\em Phys. Rev.} {\bf D93} (2016), no.~10, 101503,
\href{http://www.arXiv.org/abs/1603.04824}{{\tt arXiv:1603.04824}}.
%%CITATION = ARXIV:1603.04824;%%.

%\bibitem{Afshar:2016kjj}
H.~Afshar, D.~Grumiller, W.~Merbis, A.~Perez, D.~Tempo, and R.~Troncoso,
  ``{Soft hairy horizons in three spacetime dimensions},'' {\em Phys. Rev.}
  {\bf D95} (2017), no.~10, 106005,
\href{http://www.arXiv.org/abs/1611.09783}{{\tt arXiv:1611.09783}}.
%%CITATION = ARXIV:1611.09783;%%.

\bibitem{Afshar:2016uax}
H.~Afshar, D.~Grumiller, and M.~M. Sheikh-Jabbari, ``{Near horizon soft hair as
  microstates of three dimensional black holes},'' {\em Phys. Rev.} {\bf D96}
  (2017), no.~8, 084032,
\href{http://www.arXiv.org/abs/1607.00009}{{\tt arXiv:1607.00009}}.
%%CITATION = ARXIV:1607.00009;%%.

%\bibitem{Sheikh-Jabbari:2016npa}
M.~M. Sheikh-Jabbari and H.~Yavartanoo, ``{Horizon Fluffs: Near Horizon Soft
  Hairs as Microstates of Generic AdS3 Black Holes},'' {\em Phys. Rev.} {\bf
  D95} (2017), no.~4, 044007,
\href{http://www.arXiv.org/abs/1608.01293}{{\tt arXiv:1608.01293}}.
%%CITATION = ARXIV:1608.01293;%%.

%\bibitem{Afshar:2017okz}
H.~Afshar, D.~Grumiller, M.~M. Sheikh-Jabbari, and H.~Yavartanoo, ``{Horizon
  fluff, semi-classical black hole microstates --- Log-corrections to BTZ
  entropy and black hole/particle correspondence},'' {\em JHEP} {\bf 08} (2017)
  087,
\href{http://www.arXiv.org/abs/1705.06257}{{\tt arXiv:1705.06257}}.
%%CITATION = ARXIV:1705.06257;%%.

%\bibitem{Hajian:2017mrf}
K.~Hajian, M.~M. Sheikh-Jabbari, and H.~Yavartanoo, ``{Fluffing Extreme
  Kerr},''
\href{http://www.arXiv.org/abs/1708.06378}{{\tt arXiv:1708.06378}}.
%%CITATION = ARXIV:1708.06378;%%.

\bibitem{essay2018}
D.~Grumiller, A.~Perez, M.~Sheikh-Jabbari, and R.~Troncoso, ``{Soft hair on
  black holes and cosmological horizons in any dimension}.'' preprint
  TUW-18-03.

\bibitem{Ashtekar:1986yd}
A.~Ashtekar, ``New variables for classical and quantum gravity,'' {\em Phys.
  Rev. Lett.} {\bf 57} (1986)
2244--2247.
%%CITATION = PRLTA,57,2244;%%.

\bibitem{Wald:1993nt}
R.~M. Wald, ``Black hole entropy is the {N}oether charge,'' {\em Phys. Rev.}
  {\bf D48} (1993) 3427--3431,
\href{http://arXiv.org/abs/gr-qc/9307038}{{\tt gr-qc/9307038}}.
%%CITATION = GR-QC 9307038;%%.
%
%\bibitem{Iyer:1994ys}

V.~Iyer and R.~M. Wald, ``Some properties of {N}oether charge and a proposal
  for dynamical black hole entropy,'' {\em Phys. Rev.} {\bf D50} (1994)
  846--864,
\href{http://arXiv.org/abs/gr-qc/9403028}{{\tt gr-qc/9403028}}.
%%CITATION = GR-QC 9403028;%%.

\bibitem{deWit:1988wri}
J.~Hoppe, ``{Quantum theory of a massless relativistic surface and a two-dimensional bound state problem},''
%J.~Hoppe, ``{Membranes and matrix models},''
\href{http://hdl.handle.net/1721.1/15717}{{MIT PhD Thesis, 1982}}.

B.~de~Wit, J.~Hoppe, and H.~Nicolai, ``{On the Quantum Mechanics of
  Supermembranes},'' {\em Nucl. Phys.} {\bf B305} (1988)
545.
%%CITATION = NUPHA,B305,545;%%.
%
%\bibitem{Hoppe:2002km}

%%CITATION = HEP-TH/0206192;%%.

%\bibitem{Jacobson:1994iw}
%\bibitem{Susskind:1994sm}
%L.~Susskind and J.~Uglum, ``{Black hole entropy in canonical quantum gravity
%  and superstring theory},'' {\em Phys. Rev.} {\bf D50} (1994) 2700--2711,
%\href{http://www.arXiv.org/abs/hep-th/9401070}{{\tt hep-th/9401070}}.
%%CITATION = HEP-TH/9401070;%%.
%
%T.~Jacobson, ``{Black hole entropy and induced gravity},''
%\href{http://www.arXiv.org/abs/gr-qc/9404039}{{\tt gr-qc/9404039}}.
%%CITATION = GR-QC/9404039;%%.

\bibitem{Ryu:2006ef}
S.~Ryu and T.~Takayanagi, ``{Aspects of Holographic Entanglement Entropy},''
  {\em JHEP} {\bf 0608} (2006) 045,
\href{http://www.arXiv.org/abs/hep-th/0605073}{{\tt hep-th/0605073}}.
%%CITATION = HEP-TH/0605073;%%.

\end{thebibliography}

\providecommand{\href}[2]{#2}\begingroup\raggedright\endgroup

\end{document}